\documentstyle[prl,aps]{revtex}
\begin{document}
\draft

\title{Laser Cooling of two trapped ions: \\
 Sideband cooling beyond the Lamb--Dicke limit.}
\author{G. Morigi$^{(1)}$, J. Eschner$^{(2)}$, J.I. Cirac$^{(1)}$
 and P. Zoller$^{(1)}$}
\address{
$^{(1)}$Institut f\"ur Theoretische Physik, Universit\"at Innsbruck, 
A--6020 Innsbruck, AUSTRIA,\\
$^{(2)}$Institut f\"ur Experimentalphysik, Universit\"at Innsbruck, 
A--6020 Innsbruck, AUSTRIA.
}
\date{\today}

\maketitle

\abstract
We study laser cooling of two ions that are trapped in a harmonic
potential and interact by Coulomb repulsion. Sideband cooling in the
Lamb--Dicke regime is shown to work analogously to sideband cooling of
a single ion. Outside the Lamb--Dicke regime, the incommensurable
frequencies of the two vibrational modes result in a quasi-continuous
energy spectrum that significantly alters the cooling dynamics. The
cooling time decreases nonlinearly with the linewidth of the cooling
transition, and the effect of trapping states which may slow down the
cooling is considerably reduced. We show that cooling to
the ground state is possible also outside the Lamb-Dicke regime.
We develop the model and use Quantum Monte Carlo calculations for
specific examples. We show that a rate equation treatment is a good
approximation in all cases.

\section{Introduction}

The emergence of schemes that utilize trapped ions or atoms for quantum
information, and the interest in quantum statistics of ultra cold atoms, have
provided renewed interest and applications for laser cooling techniques
\cite{Reviews}. The present goal is to laser cool several atoms to a pure
quantum state (to the motional ground state), and experimental \cite{Salomon}
and theoretical efforts \cite{Lewenstein} are made in this direction. The
cooling of a large number of particles using lasers is a prerequisite for
coherent control of atomic systems \cite{BED,Zoller}.  In quantum information,
for example, laser cooling to the motional ground state is a fundamental step
in the preparation of trapped ions for quantum logic \cite{Zoller}. 
Coherent control and
manipulation of information requires that each ion be individually addressable
with a laser, and thus restricts the choice of the trap frequency, and
consequently the regime in which cooling
must work, to relatively shallow traps. \\
Laser cooling of single ions in traps has been extensively studied
\cite{Stenholm} and in particular sideband cooling has been demonstrated to be
a successful technique for cooling single ions to the ground state of a
harmonic trap \cite{Monroe}.  Sufficient conditions for sideband cooling a two
level system are: {\it (i)} the radiative linewidth $\gamma$ is smaller than
the trap frequency $\nu$, such that motional sidebands, i.e.  optical
transitions that involve the creation or annihilation of a specific number of
motional quanta, can be selectively excited; {\it (ii)} the Lamb--Dicke limit
is fulfilled, i.e. the ion's motional excursion is much smaller than the laser
wavelength.  The first condition can be achieved through an adequate choice of
atomic transition or a manipulation of the internal atomic structure
\cite{Irene}, while the Lamb--Dicke regime requires the trap frequency to be
much larger than the recoil frequency of the optical
transition. \\
For more than one ion, as required in quantum logic schemes, individual
addressing imposes small trap frequencies, whereas sideband cooling imposes
high trap frequencies.  Furthermore, the Coulomb interaction between the
particles makes the problem much more complex, and it is not obvious whether
the techniques developed for single ions can be transferred directly to this
situation.  Experimentally, sideband cooling of two ions to the ground state
has been achieved in a Paul trap that operates in the Lamb--Dicke limit
\cite{Wineland}.  However in this experiment the Lamb--Dicke regime required
such a high trap frequency that the distance between the ions does not allow
their individual addressing with a laser.  For this purpose, and also for an
extension beyond two ions, linear ion traps \cite{Linear} are the most
suitable systems, and for their physical parameters laser cooling to the
ground state is a goal yet to be achieved. Laser cooling of two ions into the
ground state is the problem that we
address in this paper. \\
Theoretical studies on cooling of single ions outside the Lamb--Dicke regime
exist \cite{noLDLio,noLDL}, while laser cooling of more than one ion has been
analyzed only in the Lamb--Dicke regime \cite{Vogt,Java}.  In this paper, we
investigate laser cooling, as developed for single ions, when it is applied to
two or more ions.  We focus our attention to sideband cooling, showing and
discussing new physical effects which arise because of the presence of two
interacting particles.  Doppler cooling will be discussed in a future work.
We show that cooling of two ions outside the Lamb--Dicke regime presents novel
features with respect to single ion cooling, and we show how the preparation
of the two ions in a pure quantum state is possible.  The results will allow
us to get some insight into the more general problem of cooling a string of
$N$ ions.  This is not only relevant for quantum logic with $N$--ion strings
but also for laser cooling of ion clusters in Paul and Penning traps
\cite{Linear,clusters}.  \\
The paper is organized as follows. In section II we introduce and discuss the
model which we will use throughout the paper, and discuss some concepts
developed in sideband cooling of one ion in relation to the presence of more
than one ion. In section III we study and discuss sideband cooling of two ions
inside and outside the Lamb--Dicke regime, and compare the two different
behaviours. Finally in the conclusions we summarize the main results, and
discuss the problem of cooling $N>2$ ions.\\

\section{Model}

We consider two ions of mass $m$ and charge $e$ placed in a one--dimensional
harmonic potential of frequency $\nu$. We assume the ions to be strongly
trapped in the other spatial dimensions so that their motion in those
directions is frozen out. Their internal structure is described by a two level
system with ground state $|\text{g}\rangle$, excited state $|\text{e}\rangle$
and resonance frequency $\omega_0$. The ions interact with laser light at
frequency $\omega_L$ and wavevector $k$. For classical laser light and in the
Rotating Wave Approximation the Hamiltonian of the system is:

\begin{equation}
H = H_i + H_{\text{mec}} + V.  \label{Hamiltonian}
\end{equation}
Here $H_i$ is the internal energy in the rotating frame
\begin{equation}
H_i=-\delta\sum_{j=1,2}|\text{e}\rangle_j{~}_j\langle \text{e}|
\end{equation}
where  $\delta=\omega_L-\omega_0$ is the detuning, $j$ labels the ion
($j=1,2$) and we have taken $\hbar=1$. 
$H_{\text{mec}}$ is the mechanical Hamiltonian,
\begin{equation}
H_{\text{mec}}=\frac{p_1^2}{2m}+\frac{p_2^2}{2m}+\frac{1}{2}m\nu^2 x_1^2 + 
\frac{1}{2}m\nu^2 x_2^2 + \frac{e^2}{4\pi\epsilon_0|x_1-x_2|}  \label{mec}
\end{equation}
with $x_j$, $p_j$ position and momentum of the $j$th ion ($j=1,2$), and 
$V$ describes the interaction between laser and atoms,
\begin{equation}
V=\sum_{j=1,2}\frac{\Omega(x_j)}{2}[\sigma_j^{\dagger}\text{e}^{-ik\cos\theta x_j} +%
\text{h.c.}].  \label{Vpot}
\end{equation}
Here $\Omega(x_j)$ is the Rabi frequency at the position $x_j$,
$\sigma_j^{\dagger}$, $\sigma_j$ are the raising and lowering dipole operators
respectively defined on the $j$th ion ($j=1,2$), and $\theta$ is the angle
between the laser wavevector and the trap axis.\\
Using the center of
mass (COM) and relative coordinates, the mechanical Hamiltonian in (\ref{mec})
is composed of two separate terms: one for the COM motion which describes a
particle of mass $M=2m$ interacting with a harmonic potential of frequency
$\nu$; the other for the relative motion which describes a particle of mass
$\mu=m/2$ interacting with a potential, which is the sum of a harmonic
potential of frequency $\nu$ and a Coulomb-type central potential. This
potential may be approximated by a harmonic oscillator potential of frequency
$\nu_r=\sqrt{3}\nu$, obtained through the truncation at the second order of
its Taylor expansion around the equilibrium distance 
$x_0=(2e^2/4\pi\epsilon_0M\nu)^{1/3}$
between the ions\cite{Vogt,Java}. In appendix A we discuss this approximation
and we show that it is valid in the regime that we are going to study. With
this approximation the term (\ref{mec}) becomes (apart for a constant)

\begin{equation}
H_{\text{mec}}=\frac{P^2}{2M}+\frac{1}{2}M\nu^2X^2+
\frac{p^2}{2\mu}+\frac{1}{2} \mu\nu_r^2x^2,  \label{new_mec}
\end{equation}
where $X=(x_1+x_2)/2$, $P=p_1+p_2$ are position and momentum of the COM
respectively, and $x=x_1-x_2-x_0$, $p=(p_1-p_2)/2$ are position and momentum
of the relative motion. Thus the term (\ref{new_mec}), apart for a constant,
can be rewritten as
\begin{equation}
H_{\text{mec}}=\nu a_0^{\dagger}a_0+\nu_r a_r^{\dagger}a_r,
\end{equation}
where we have defined $X=\sqrt{1/2M\nu_0}(a_0^{\dagger}+a_0)$,
$P=i\sqrt{M\nu_0/2}(a_0^{\dagger}-a_0)$,
$x=x_0+\sqrt{1/2\mu\nu_r}(a_r^{\dagger}+a_r)$, $p=i\sqrt{
  \mu\nu_r/2}(a_r^{\dagger}-a_r)$, with $a_0$, $a_0^{\dagger}$ the
annihilation and creation operators for the COM mode respectively, and $a_r$,
$a_r^{\dagger}$ the corresponding ones for the relative motion (stretch) mode.
We stress that in this new representation the mechanical problem of two ions
interacting through Coulomb forces is reduced to the one of two harmonic
oscillators, while the interaction of each ion with the radiation is now
transformed into a nonlinear coupling between the harmonic oscillators. 
In general, $N$ ions in a trap can be described by a set of $N$ harmonic
oscillators, coupled by laser light \cite{James}.\\
The master equation for the density matrix $\rho$ of the two ion system is  
\cite{Vogt}

\begin{equation}
\frac{\text{d}}{\text{d}t}\rho
=-\frac{i}{\hbar}[H,\rho]+L\rho  \label{Master}.
\end{equation}
Here $L$ is the Liouvillian describing the incoherent evolution of the system:

\begin{equation}
L\rho=\frac{\gamma}{2} \sum_{j=1,2} 
[2\sigma_j\tilde{\rho}_j\sigma_j^{\dagger}
-\sigma_j^{\dagger} \sigma_j\rho-\rho\sigma_j^{\dagger}\sigma_j],
\end{equation}
where $\gamma$ is the decay rate out of the internal excited state
$|\text{e}\rangle$, and $\tilde{\rho}_j$ describes the density matrix after a
spontaneous emission for the $j$th ion:

\begin{equation}
\label{feed}
\tilde{\rho}_j=\int_{-1}^1\text{d}uN(u)\text{e}^{ikux_j}\rho\text{e}^{-ikux_j}
\end{equation}
with $N(u)$ being the dipole pattern for the decay.\\
In this treatment we have neglected both dipole--dipole interaction between
the ions and quantum statistical properties. This approximation is justified
in the regime that we investigate, which is characteristic of experiments
using linear ion traps for quantum information \cite{Linear}. In those traps
the equilibrium distance between the ions is of the order of 10$\mu$, while
the laser wavelength is typically in the visible region and the individual
ionic wavepackets have spatial widths of the order of 10-100 nm.  From these
considerations we can consider the two ions in a linear trap as two
distinguishable particles in a harmonic potential which interact solely with
Coulomb forces. 
On the basis of these considerations we will use Eq. (\ref{Master})
for the numerical simulations presented below.\\
For the following discussion it is instructive to look at the set of
equations which one obtains from (\ref{Master}) in the limit of low
saturation $\Omega\ll \gamma$, when the excited state $|\text{e}\rangle$ can
be eliminated in second order perturbation theory (we provide a detailed
derivation of the equations in Appendix B). In the basis of states
$|\text{g},{\bf n}\rangle$, where ${\bf n}=(n_0,n_r)$ is a vector with COM
vibrational number $n_0$ and stretch mode vibrational number $n_r$, we have
the following set of equations:

\begin{eqnarray}
\label{no_neglect}
\frac{d}{dt}\langle \text{g},{\bf n}|\rho |\text{g},{\bf m}\rangle 
&=&-i({\bf n}-{\bf m})\cdot 
\upsilon\langle \text{g},{\bf n}|\rho |\text{g},{\bf m}\rangle  
\label{Equa} \\
&+&i\frac{\Omega ^{2}}{4}\sum_{{\bf k},{\bf l}}\left[ 
\frac{\langle{\bf n}|\text{e}^{ik\cos\theta x_1}|{\bf k}\rangle 
\langle {\bf k}|\text{e}^{-ik\cos\theta x_1}|{\bf l}\rangle }
{({\bf k}-{\bf l})\cdot \upsilon-\delta -i\gamma/2 }
\langle \text{g},{\bf l}|\rho |\text{g},{\bf m}\rangle 
-\langle \text{g},{\bf n}|\rho |\text{g},{\bf l}\rangle 
\frac{\langle {\bf l}|\text{e}^{ik\cos\theta x_1}|{\bf k}\rangle 
\langle {\bf k}|\text{e}^{-ik\cos\theta x_1}|{\bf m}\rangle }
{({\bf k}-{\bf l})\cdot \upsilon-\delta +i\gamma/2}\right] 
\nonumber\\
&+&\frac{\Omega ^{2}}{4}\sum_{{\bf k},{\bf j},{\bf r},{\bf s}}
\int_{-1}^{1}\text{d}uN(u)
\langle {\bf n}|\text{e}^{ikux_1}|{\bf k}\rangle 
\langle {\bf k}|\text{e}^{-ik\cos\theta x_1}|{\bf r}\rangle 
\langle {\bf s}|\text{e}^{ik\cos\theta x_1}|{\bf j}\rangle 
\langle {\bf j}|\text{e}^{-ikux_1}|{\bf m}\rangle 
\langle \text{g},{\bf r}|\rho |\text{g},{\bf s}\rangle
\nonumber \\
&\cdot &\Bigl[\frac{1}{\left( ({\bf j}-{\bf s}-{\bf k}+{\bf r})\cdot 
\upsilon+i\gamma \right) 
\left( ({\bf j}-{\bf s})\cdot \upsilon-\delta-i\gamma/2 \right) }
+\frac{1}{\left( ({\bf k}-{\bf r}-{\bf j}+{\bf s})\cdot 
\upsilon-i\gamma \right) \left( ({\bf k}-{\bf r})\cdot 
\upsilon-\delta+i\gamma/2 \right) }\Bigr],
\nonumber
\end{eqnarray}
where $\upsilon=(\nu ,\nu_r)$ and where for simplicity we have assumed that
only ion 1 is illuminated, i.e. $\Omega (x_{1})=\Omega$, $\Omega (x_{2})=0$.
This implies that we can address the ions individually with a well focused
laser beam. It corresponds, for example, to a situation where the two ions are
two different isotopes, of which only one is resonant
with light \cite{Lange}.\\
When treating laser cooling in a harmonic trap, an important dimensionless
quantity is the Lamb--Dicke parameter $\eta$, which for a single ion of mass
$m$ in a trap of frequency $\nu$, interacting with laser light of wavevector
$k$ is:

\begin{equation}
\label{LD_0}
\eta=k\sqrt{\frac{1}{2m\nu}}=\sqrt{\frac{\omega_{\text{rec}}}{\nu}},
\end{equation}
where $\omega_{\text{rec}}=k^2/2m$ is the recoil frequency.  The parameter
$\eta$ appears in the kick operator $\exp(ikx)$ in the term describing the
exchange of momentum between radiation and atoms, which using the relation
$x=\sqrt{1/2m\nu}(a^{\dagger}+a)$ and the definition (\ref{LD_0}) is rewritten
as $\exp(ikx)=\exp(i\eta(a^{\dagger}+a))$.  The
Lamb--Dicke regime corresponds to the condition $\sqrt{n}\eta\ll 1$, with $n$
vibrational number; in other words, to the situation in which, during a
spontaneous emission, a change in the vibrational number of the atomic state is
unlikely due to energy conservation. In this regime the kick operator may be
expanded in powers of $\eta$, and with good approximation the expansion may be
truncated at the first order \cite{Stenholm}.
Another important parameter, as known from cooling of single ions, is the
ratio between the radiative linewidth $\gamma$ and the trap frequency $\nu$:
in the so--called strong confinement regime $\gamma/\nu\ll 1$ the laser can
selectively excite sidebands of the optical transition which involve a
well--defined change of the vibrational number $n$.  In this regime, together
with the Lamb--Dicke regime, sideband cooling works efficiently: when the
laser is red detuned with $\delta=-\nu$ the system is cooled by approximately
one phonon of energy $\nu$ in each fluorescence cycle, finally reaching the
vibrational
ground state $n=0$ \cite{Monroe}.
In contrast, in the weak confinement regime $\gamma/\nu \ge 1$, transitions
which involve different changes of the vibrational number $n$ are excited
simultaneously.  This is the Doppler cooling regime, where the achievable
minimum energy for a single ion is approximately $\gamma/2$ for a detuning
$\delta=-\gamma/2$ \cite{footnote1}.\\

Having introduced these basic concepts and methods of laser cooling of single
ions in traps, we now turn back to the problem of two ions, to discuss how
those techniques may be applied, and whether the same concepts are still
valid.  For $N$ harmonic oscillator modes we can define a Lamb--Dicke
parameter for each mode in an analogous way to (\ref{LD_0}). 
For our case, $N=2$,
the Lamb--Dicke parameters $\eta_0$ for the COM mode and $\eta_r$ for the
stretch mode are defined as:

\begin{eqnarray}
\label{LD}
\eta_0&=&k\sqrt{\frac{\hbar}{2M\nu}}=\frac{\eta}{\sqrt{2}},\\
\eta_r&=&
\frac{k}{2}\sqrt{\frac{\hbar}{2\mu\nu_r}}=\frac{\eta}{\sqrt{2\sqrt{3}}},
\nonumber
\end{eqnarray}
so that the kick operator for the $j$th ion ($j=1,2$) is written as:
\begin{equation}
\text{e}^{ikx_j}=\text{e}^{i\eta_0(a_0^{\dagger}+a_0)}
\text{e}^{(-1)^{j-1} i\eta_r(a_r^{\dagger}+a_r)}.
\label{kick}
\end{equation}
In general for $N$ ions $\eta_{0}=\eta/\sqrt{N}$ \cite{James}. In the
following, when we refer to the Lamb--Dicke regime we will consider the
situation where the conditions $\sqrt{n_0}\eta_0\ll 1$, $\sqrt{n_r}\eta_r\ll
1$ are fulfilled. From (\ref{kick}) we see that in Eq. (\ref{Vpot}) the
Lamb--Dicke parameters appear multiplied by the factor $\cos\theta$.
Therefore the Lamb--Dicke parameters for the coherent excitation,
$\eta_i\cos\theta$, are always less than or equal to the Lamb--Dicke
parameters
defined in (\ref{LD}), which characterize the spontaneous emission.\\
To discuss the importance of the ratio $\gamma/\nu$ in the case of two
ions, we first consider the bare spectrum of energies of our system with
frequencies $\nu$ and $\sqrt{3}\nu$.  In Fig. 1 we plot the density of states
$D(E)$ {\it vs}. the energy $E$, defined as the number of states in the
interval of energy $[E,E+\delta E]$.
From this figure we see that due to the {\it incommensurate} character of the
frequencies the spectrum does not exhibit a well distinguished series of
energy levels, rather it tends to a quasi--continuum. Therefore, the strong
confinement requirement for sideband cooling needs to be reconsidered. The
main question which we will address in the following is whether it is still
possible to cool one mode to the ground state by means of sideband cooling. As
we will show, the Lamb--Dicke parameter distinguishes two regimes which
exhibit dramatic differences.

\section{Sideband cooling of two ions}

In the following we study sideband cooling of two ions, first in the
Lamb--Dicke regime and then outside of this regime. We will show that in this
latter case two--ion effects appear due to the dense spectrum of energy
levels.  In our calculations we first consider sideband cooling when laser
light excites only one of the two ions directly. Thus in (\ref{Vpot}) we take
$\Omega(x_1)=\Omega$, and $\Omega(x_2)=0$. Afterwards we compare this case to
the one in which both are driven by light, i.e.
$\Omega(x_1)=\Omega(x_2)=\Omega$, showing that the only difference between the
two cases is the cooling time, which in the latter case scales by a factor
1/2. In the following we assume the laser wavevector parallel to the trap
axis, i.e. $\cos\theta=1$. This assumption facilitates the analysis and it is
justified by the simple scaling just described. Furthermore it corresponds to
the case in which the two ions are two different ionic isotopes, of which one
is driven by light \cite{Lange}. At the end of this section we will briefly
discuss cooling of two identical ions when the wavevector is not parallel to
the trap axis.

\subsection{Lamb--Dicke regime.}

In the Lamb--Dicke regime the Franck--Condon coefficients $\langle {\bf
  n}|\exp(ikx)|{\bf l}\rangle$ in the numerators of the right hand side terms
of Eq. (\ref{no_neglect}) may be expanded in terms of the Lamb--Dicke
  parameters $\eta_0$, $\eta_r$.
The response of the system to laser light is governed by its spectrum of
resonances $I(\delta)$, which is evaluated by summing all contributions to
laser--excited transitions at frequency $\omega_L$,
\begin{equation}
\label{reson}
I(\delta)=\sum_{({\bf n}-{\bf l})\cdot \upsilon=\delta}
|\langle {\bf n}|\exp(ikx)|{\bf l}\rangle|^2 
P({\bf n})
\end{equation}
where $P({\bf n})$ is a normalized distribution of the states $|{\bf
  n}\rangle$.  In
the Lamb--Dicke regime, we find that $I(\delta)$ exhibits two main pairs of
sidebands around the optical frequency $\omega_0$: one at frequencies
$\omega_0\pm \nu_0$ corresponding to the transition $n_0\to n_0\pm 1$; the
other at frequencies $\omega_0\pm \nu_r$ corresponding to $n_r\to n_r\pm 1$
(see Fig. 2(a)). The strength of these sidebands relative to the carrier ${\bf
  n}\to {\bf n}$ are proportional to $\eta_0^2$ and to $\eta_r^2$,
respectively. All the other sidebands have strengths of higher orders in
$\eta_0^2$, $\eta_r^2$.  This implies that by selecting one of these four
sidebands by laser excitation we will induce the corresponding phononic
transition; for example by choosing the sideband corresponding to
$n_0,n_r\to n_0-1,n_r$ we can cool the COM mode to its vibrational ground
state, as for a single ion.  This has been experimentally demonstrated by the
NIST group at Boulder \cite{Wineland}.  In Fig. 2 we plot the results of a
Quantum Monte Carlo wavefunction simulation (QMC) of Eq. (\ref{Master}) for two
ions in a trap with Lamb--Dicke parameter $\eta_{0}=0.1$, radiative linewidth
$\gamma=0.2\nu$, detuning $\delta=-\nu$ and an initially flat distribution for
the states with energy $E\le 15\nu$. In Fig. 2(b) the average vibrational
numbers of the COM mode (solid line) and of the stretch mode (dashed line) are
plotted as a function of time in unit of fluorescence cycles
$t_F=2\gamma/\Omega^2$.  The system behaves as if the two modes were
decoupled, since only one mode is cooled while the other remains almost
frozen. Nevertheless the stretch mode is cooled on a much longer time scale,
as an effect of off--resonant excitation.  In Fig. 2(c) the populations of the
vibrational states of the two modes are plotted at time $t=600t_F$, showing
the COM mode in the ground state and the nearly uncooled stretch mode.
In this limit we can neglect the coupling of the population to the coherences
in Eq. (\ref{no_neglect}), thus reducing the density matrix equation in the low
saturation limit to rate equations. In fact the
coherences in Eq. (\ref{no_neglect}) have either an oscillation frequency
which is much larger than the fluorescence rate $1/t_F$, or a coupling to the
population which is of higher order in the Lamb--Dicke parameter expansion, or
both of them. Therefore, they can be neglected in the equations of the
populations, and we obtain the set of rate equations:

\begin{eqnarray}
\frac{d}{dt}\langle {\bf n}|\rho |{\bf n}\rangle &=&-\gamma 
\frac{\Omega^{2}}{4}
\sum_{{\bf k}}\frac{|\langle {\bf n}|\text{e}^{ikx_1}|{\bf k}\rangle |^{2}}
{\left[ ({\bf k}-{\bf n})\cdot\upsilon-\delta \right] ^{2}
+\gamma ^{2}/4}
\langle {\bf n}|\rho |{\bf n}\rangle  \nonumber \\
&+&\gamma \frac{\Omega ^{2}}{4}\sum_{{\bf k},{\bf r}}
\int_{-1}^{1}\text{d}uN(u)
\frac{|\langle {\bf n}|\text{e}^{ikux_1}|{\bf k}\rangle|^{2}|\langle {\bf k}|
\text{e}^{-ikx_1}|{\bf r}\rangle |^{2}}{\left[ ({\bf k}-{\bf r})\cdot 
\upsilon
-\delta \right] ^{2}+\gamma ^{2}/4}\langle {\bf r}|\rho |{\bf r}\rangle
\label{rate}
\end{eqnarray}
where we have omitted the label $\text{g}$ of the states.
The validity of this approximation is shown for the above case in Fig. 3(a),
where the results of figure 2(b) are compared with those of a rate
equation simulation according to Eq. (\ref{rate}).
As a further proof that the two modes can be considered decoupled during the
time in which the COM motion is cooled, in Fig. 3(b) we compare the time
dependence of the average vibrational number of the COM mode with the one of a
single trapped ion which is cooled under the same Lamb--Dicke parameter,
radiative linewidth, trap frequency, Rabi frequency and initial distribution
as the COM mode. We see that the two curves overlap appreciably, justifying the
picture of sideband cooling of two ions in the Lamb--Dicke regime as if the
modes were decoupled from one another.

\subsection{Outside the Lamb--Dicke regime}

To illustrate the physical features of the system outside the Lamb--Dicke
regime, we plot in Fig. 4(a) the spectrum of resonances $I(\delta)$ as defined
in Eq. (\ref{reson}) for two ions in a harmonic trap with $\eta_0=0.6$.  We
see that the spectrum exhibits many sidebands whose density increases as the
detuning increases. The main consequence is that we cannot select a given
sideband by choosing the laser frequency, but rather excite a group of
resonances that correspond to transitions to a set of quasi--degenerate
states.  The range of transitions that are excited increases with $\gamma$.
In figure 4(b),(c) we consider sideband cooling for Lamb--Dicke parameter
$\eta_0=0.6$ and detuning $\delta=-2\nu$ \cite{noLDLio}, where the other
parameters are the same as in Fig. 2(b),(c).
As one can see, the two modes are coupled and cooled together. Thus, as a
first big difference with respect to single ion cooling, we see that here the
energy is not taken away from one mode only, rather it is subtracted from the
system as a whole. Another striking difference appears in the cooling time,
which is significantly longer in comparison with the time necessary to cool
one single ion outside the Lamb--Dicke regime\cite{noLDLio}. This slowing down
is partly due to the increase of the dimension of the phase space where the
cooling takes place: the presence of two modes makes the problem analogous to
cooling in a two dimensional trap, whose axis are coupled by the laser. The
ions thus make a random walk in a larger phase space, and the cooling gets
slower.  However the cooling time is even considerably longer than one would
expect taking the dimensionality into account. This can be explained by
looking again at the spectrum of resonances in Fig. 4(a): despite the high
density of resonances, the coupling between the states is still governed by
the Franck--Condon coupling, i.e.  by the terms in the numerator of Eq.
(\ref{no_neglect}) which outside the Lamb--Dicke limit oscillate with the
vibrational numbers of the states. In the limit of linewidth $\gamma\ll\nu$,
where a single sideband can be selected, we may encounter {\it trapping
states} like in cooling of single ions \cite{Parkins}, i.e. states whose
coupling to the resonantly excited state is very small since their motional
wavefunction after the absorption of a laser photon happens to have a very
small overlap with the motional wavefunction of the excited states. This
effect limits the cooling efficiency, since the atoms may remain trapped in
these states and not be cooled further, or much more slowly, towards the
ground state. For two ions the probability of finding zeroes of the Franck
Condon coupling is larger than for one ion, as the coupling to the excited
state is constituted by two integrals, one for the COM and the other for the
relative motion wavefunctions. Thus the probability of having trapping states
is higher. To illustrate this phenomenon we plot in Fig. 5(a) the occupation
of the states $P_{\bf n}$ as a function of the COM and relative vibrational
numbers $n_0$ and $n_r$, respectively, at a time $t=1000t_F$ after sideband
cooling of the COM with $\delta=-2\nu$. Here $\gamma=0.02\nu$, and we are in
the limit in which the single resonances are resolved. As a consequence the
most likely coherent transitions are $n_0\to n_0-2$, $n_r\to n_r$. The effect
of the trapping states is visible in the tail of occupied states of $P_{\bf
  n}$, with $n_r=6,7$. In Fig. 5(b),(c) we plot the modulus square of the
Franck Condon coefficients for the relative motion corresponding to the
coupling of the states $n_r=6,7$ to the other motional states respectively:
here it is clearly shown that for the transition $n_r=6\to 6$, $n_r=7\to 7$, 
the coupling is reduced nearly to zero.
As the linewidth $\gamma$ increases, the number of states to which a single
state is coupled increases. Thus the number of channels through which the atom
may be cooled is larger. As an effect the trapping states disappear. This is
shown in Fig. 6, where the population $P_{\bf n}$ is plotted for $t=600t_F$
and $\gamma=\nu$, and otherwise the same parameters as before. Here we see
that the system is cooled homogeneously.
The effect of varying $\gamma$ is summarized in Fig. 7, where we compare the
average COM vibrational number {\it vs}. time in unit $t_F$ for various values
of $\gamma$.
The results of Fig. 7 show clearly that as the linewidth increases the number
of fluorescence cycles needed for cooling the system decreases dramatically.
It is important to note that in this diagram the time is measured in units of
fluorescence cycles for each $\gamma$, so that the absolute cooling time
clearly reduces more strongly. We stress that this strong dependence on the
linewidth is a two--ion effect. In contrast, in sideband cooling of single
ions, the fluorescence time $t_F$ determines the cooling time scale for
$\gamma/\nu\le 1$, and the curves for different values of $\gamma$ {\it vs.}
the time in unit of the
respective $t_F$ do not show striking differences.\\
The presence of trapping states and the coupling of each state to more than
one state at almost the same transition frequency might lead to the formation
of {\it dark coherences} between quasi--degenerate states, i.e. to
superpositions of states which decouple from laser excitation because of
quantum interference.  However for the considered system those dark states do
not play any significant role. We prove this numerically in Fig. 8, where we
plot the comparison between a QMC and a rate equation simulation. We see that
there are no striking differences between the two curves. We point out that
outside the Lamb--Dicke regime a rate equations treatment is not justified in
principle, since secular approximation arguments and Lamb--Dicke limit
arguments cannot be applied.  Here the rate equations are used to highlight
the effect of neglecting the coherences in the dynamical evolution of the
cooling, while these coherences are fully accounted for in the QMC treatment.
In order to see why coherences do not play any significant role in the cooling
dynamics, we look at the definition of a dark state.  Let us consider 
a state $|\alpha\rangle$ at $t=0$
defined for simplicity as linear superposition of two
quasi--degenerate states $|\alpha\rangle=a_1|{\bf
  n}\rangle+a_2\text{e}^{i\phi}|{\bf m}\rangle$, with $a_1$, $a_2$, $\phi$
real coefficients and with ${\bf n}$, ${\bf m}$ states almost degenerate in
energy, so that $|({\bf n}-{\bf m})\cdot \upsilon| =\Delta E$ with $\Delta E
\le\gamma$. In principle a dark state can be a linear superposition of any
number of states. However as we will see from the arguments below our
restriction to two states does not affect the generality of the result. The
evolution $|\alpha(t)\rangle$ in the Schroedinger picture, apart from a global
phase factor, is written as
\begin{equation}
|\alpha(t)\rangle=a_1|{\bf n}\rangle+a_2\text{e}^{i\phi(t)}|{\bf
  m}\rangle,
\label{darksta}
\end{equation}
with $\phi(t)=\phi_0+\Delta E t$. The state is dark when the following 
condition is fulfilled: 
\begin{equation}
\langle {\bf  l}|\text{e}^{ik(X+x/2)}|\alpha(t)\rangle\sim 0 
\label{dark}
\end{equation}
for any state $|{\bf l}\rangle$ belonging to the set of states $\{|{\bf
  l}\rangle\}$ to which it is resonantly or almost resonantly coupled.  If the
condition (\ref{dark}) holds at $t=0$, it will hold up to a time $t$ such that
$\Delta E t\sim \pi/2$.  For the system we are dealing with we do not have
exact degeneracy, thus we check whether the state $|\alpha\rangle$ can remain
dark for a time sufficiently long to affect the cooling dynamics appreciably.
The smallest possible value of $\Delta E$ in the range of energies of our
calculations is $\Delta E=0.07\nu$, and we find that $\phi$ rotates by an
angle $\pi/2$ in less than one fluorescence cycle for the values of $\gamma$
that we have considered. The dark coherences are then washed away during the
evolution as an effect of the incommensurate frequencies between the two
modes. This result together with the numerical results suggests that rate
  equations can be used in the study of cooling \cite{footnote2}. \\
In order to highlight that the absence of dark coherences is a signature of
the peculiar spectrum of the system we plot in Fig. 9 the cooling of one mode
outside the Lamb--Dicke regime for the case of a discrete spectrum where we
have exact degeneracy. More precisely we consider two harmonic oscillators with
commensurate frequencies $\nu$ and $2\nu$, where all the other physical
parameters are the same as before. In this case the different outcome between
the QMC and the rate equations treatment is dramatic, giving evidence to the
role of the coherences in the evolution.

\subsection{Light on both ions}

The calculations that we have shown refer to the case in which only one ion is
illuminated. As we have seen, although light interacts with one ion it couples
with both modes simultaneously, as shown in Eqs. (\ref{Vpot}) and
(\ref{kick}).  When both ions are excited by laser light, the system is
described by a 4--level scheme, corresponding to the 4 internal states
$|a_1,b_2\rangle$ with $a,b=e,g$, where we assume that when a photon is
emitted, we detect from which ion the event has occurred, as a consequence of
the spatial resolution of the ions.  We expect that the effect on the cooling
will be a doubling in the number of quantum jumps and hence of the cooling
rate.  This is shown in Fig. 10, where we compare the time dependence of the
COM vibrational number for the cases in which only one ion is illuminated
(dashed line) or both ions are illuminated with the same laser intensity
(solid line with label 1). In the latter case cooling is visibly faster, and
the time dependence scales with a factor of two with respect to the case with
one ion illuminated, as we can see when we replot the solid line 1 {\it vs}.
$t/2t_F$ (solid line with index 2).
The two--ions effects found above are clearly independent of the number of
scattering points, with the only difference that the dark state condition is
now written as $\langle {\bf
  l}|\left(\text{e}^{ik(X+x/2)}+\text{e}^{ik(X-x/2)}\right)
|\alpha(t)\rangle\sim 0$. We point out again that effects due to interference
between the internal excitation paths have been neglected, as we consider
the ions to be two distinguishable particles \cite{footnote5}. \\
We would like to stress that in the above calculations we have considered the
case of only one ion driven by radiation while the laser wavevector is
parallel to the trap axis. However if one wants to cool two identical ions by
shining light on one of them the laser beam must necessarily be at a certain
angle $\theta$ with respect to the ion string. Thus the Lamb--Dicke parameter
characterizing the coherent excitation will be smaller than the Lamb--Dicke
parameter for the spontaneous decay, and depending on the minimum angle
$\theta$ required, the coherent laser will excite with some selectivity one of
the two modes. However for cooling purposes it is preferable to have the two
Lamb--Dicke parameters values, corresponding to the spontaneous emission and
the coherent excitation, as close as possible. \\

\section{Conclusions}

In this paper we have studied the question of cooling two ions in a linear
trap to the ground state by means of sideband cooling.  We have studied
sideband cooling in the Lamb--Dicke regime, and we have shown that in this
limit the two harmonic oscillators can be considered decoupled when one of the
two is cooled by means of sideband cooling. We have found that the cooling
dynamics in the low intensity limit may be described by rate equations, and
essentially that all the considerations developed for sideband cooling of
single ions apply.  This regime has been considered by Vogt et al. \cite{Vogt}
for studying the effects of dipole--dipole interaction in laser cooling of two
ions and by Javanainen in
\cite{Java} in his study on laser cooling of ion clusters.\\
We have then investigated laser cooling outside the Lamb--Dicke regime,
finding striking differences with the Lamb--Dicke limit. Here the energy
spectrum may be considered a quasi--continuum, though the coupling between the
states is still governed by the Franck--Condon coupling. A consequence is that
the cooling efficiency depends strongly on the radiative linewidth.  For very
small ratio $\gamma/\nu$ the effect of trapping states is appreciable, and
manifests itself in a drastic increase of the cooling time, i.e. the number of
fluorescence cycles needed to cool the system.  For $\gamma/\nu<1$, but large
enough, the effect of trapping states is washed away, the modes are cooled
simultaneously, and the cooling time is considerably shorter and comparable to
the time needed for cooling single ions outside of the Lamb--Dicke regime.
These effects are all consequences of the density of states in the energy
spectrum. A further property of the system is the absence of dark coherences,
as a consequence of the incommensurate frequencies of the harmonic
oscillators, i.e. of the absence of perfect degeneracy. This implies that in
the low intensity limit rate equations provide still a good description of the
cooling dynamics.  Finally we have compared cooling when light is shone on one
ion only or on both ions, finding a simple difference of a factor two in the
rate of cooling.\\
On the basis of the obtained results we would like to comment on cooling of a
string of $N>2$ ions. $N$ ions in a harmonic trap may be described by $N$
harmonic oscillators. The mode frequencies $\nu_0,\nu_1,...,\nu_{N-1}$ are all
incommensurate, and the number of states in the interval of energy
$[E,E+\delta E]$ is $D(E)=E^{N-1}/\nu_0\nu_1...\nu_{N-1}\delta E$. Outside the
Lamb--Dicke regime the spectrum of resonances is then even more dense than in
the two--ions case, and the probability of trapping states will be larger.
However a suitable increase of the linewidth will cancel their effects, since
each state will see an even higher number of states, thus of possible
transitions, than for two ions. We expect that this last property will also
play a role against the creation of dark states. In fact, although on one hand
the large density of states may favour the appearance of stable dark
coherences between ``accidentally'' degenerate states, on the other hand for
values of $\gamma$ large enough the coherent effects will wash out because of
the coupling to a ``continuum'' of states.  From these considerations we
expect laser cooling to the ground state to be still possible for $N>2$ ions.
Furthermore, one can cool a given set of modes to the ground state through the
choice of the laser detuning outside the Lamb--Dicke regime.  For example
taking a detuning $\delta=-\nu_k$ the modes with frequency $\nu_j\ge \nu_k$
may be cooled to their vibrational ground states. It should be noticed that as
the number of ions $N$ increases, the Lamb--Dicke parameter of each mode
decreases approximately as $1/\sqrt{N}$ \cite{James}, allowing to reach the
Lamb--Dicke regime also when this condition is not fulfilled for single ions.
In this limit the modes may be considered decoupled and sideband cooling
is particularly efficient.\\
As a last consideration, we note that the behaviour of a number of ions $N\ge
3$ cooled by light depends on which ions of the string are driven. In fact
each position of the string couples with the different modes with amplitudes
that depend on the position itself \cite{James}. Only if all the ions are
illuminated we may consider all the modes as coupled and cooled
simultaneously. But this ``coupling'' changes as we select and drive only
certain ions of the string. In this case a certain amount of modes may be
cooled, while the others will remain hot or get cooled on a longer time scale.
In this respect the system can be considered as having a reduced
dimensionality, and the time of the cooling will be accordingly shorter with
respect to the case in which the laser couples to all modes.

\section{Acknowledgements}
We would like to thank D. Leibfried, F. Schmidt--Kaler, H.
Baldauf and W. Lange for many stimulating discussion. One of us (G.M.)
whishes to thank S. Stenholm for stimulating discussions. This work was
supported by the Austrian Fond zur F\"orderung der wissenschaftlichen
Forschung and the TMR network ERBFMRX-CT96-0002.

\appendix

\section{Harmonic approximation}

We consider the term (\ref{mec}) and rewrite it in COM and relative motion
canonical variables:
\begin{eqnarray}
\label{H_due}
H_{\text{mec}}
&=&\frac{P^2}{2M}+\frac{1}{2}M\nu^2 X^2 +
\frac{p^2}{2\mu}+\frac{1}{2}\mu\nu^2 x^2 + 
\frac{e^2}{4\pi\epsilon_0|x|}\nonumber\\
&=&H(X,P)+H(x,p).
\end{eqnarray}
with $\mu=m/2$ reduced mass, $M=2m$ total mass. The mechanical problem is
separable into center of mass motion and relative motion, where $H(X,P)$
describes the harmonic motion of a particle of mass $M$ interacting with a
harmonic oscillator of frequency $\nu$, and $H(x,p)$ the motion of a particle
of mass $\mu$ interacting with a potential $V(x)=\mu\nu^2 x^2/2 +
e^2/4\pi\epsilon_0|x|$, i.e. a harmonic potential of frequency $\nu$ and a
central repulsive
Coulomb--type potential.\\
We focus our attention on the potential $V(x)$, restricting its domain on the
semiaxis $x>0$. The equilibrium point is found to be
$x_0=\left(e^2/4\pi\epsilon_0\mu\nu^2\right)^{1/3}$. 
Expanding $V(x)$ around $x_0$ we find:

\begin{eqnarray}
V(x)
&=&\frac{3}{2}\left(\frac{e^2}{4\pi\epsilon_0}\mu\nu^2\right)^{\frac{1}{3}}
+\frac{3}{2}\mu\nu^2(x-x_0)^2+\sum_{n=3}^{\infty}(-1)^n 
\frac{e^2}{4\pi\epsilon_0 x_0^{n+1}}(x-x_0)^n\\
&=&\frac{3}{2}\left(\frac{e^2}{4\pi\epsilon_0}\mu\nu^2\right)^{\frac{1}{3}}
+\frac{1}{2}\mu\nu_r^2(x-x_0)^2+A(x),
\end{eqnarray}
where $\nu_r=\sqrt{3}\nu$ and $A(x)$ sum over the higher order terms, which we
call the anharmonic terms. We quantize the oscillation around the equilibrium
position $x_0$;

\begin{equation}
x=x_0+\sqrt{\frac{1}{2m\nu_r}}\left(a_r^{\dagger}+a_r\right),
\end{equation}
where $a_r$, $a_r^{\dagger}$ are annihilation and creation operators
respectively. From perturbation theory we may consider $V(x)$ harmonic when
the following conditions are fulfilled:

\begin{eqnarray}
\label{conditions}
&\langle j|A(x) &|j\rangle\ll \hbar\nu_r\\
&|\langle j|A(x)&|j\pm 1\rangle|\ll \hbar\nu_r\nonumber
\end{eqnarray}
with $|j\rangle$ eigenstate of the harmonic oscillator of frequency $\nu_r$
with eigenvalue $j\hbar\nu_r$.  The first condition means that the energy
shift due to the anharmonic term is much smaller than the spectrum separation,
whereas the second condition means that the coupling between the states is a
small perturbation, and we will show that it may be neglected. The coupling
between the state $j$ and the state $j+k$ is not taken here into account for
simplicity, but it may be shown that it is much smaller than $k\hbar\nu_r$ in
a similar way to the one we discuss below. Let us rewrite the relations
in (\ref{conditions}) as:
\begin{eqnarray}
\label{serie}
&\langle j&|A(x)|j\rangle=\frac{e^2}{4\pi\epsilon_0 x_0}
\sum_{m=2}^{\infty}\zeta_{j,2m}\\
&\langle j&|A(x)|j+1\rangle=-\frac{e^2}{4\pi\epsilon_0 x_0}
\sum_{m=1}^{\infty}\zeta_{j,2m+1}
\nonumber
\end{eqnarray}
with 
\begin{equation}
\label{parameter}
\zeta_{j,2m}=\left(\sqrt{\frac{1}{x_0^22\mu\nu_r}}\right)^{2m}\langle j|
\left(a_r^{\dagger}+a_r\right)^{2m}|j \rangle,
\end{equation}
where $\zeta_{j,2m+1}$ is analogously defined. From the following relation

\begin{eqnarray}
&\frac{(j+m)!}{j!}&<
\langle j|\left(a_r^{\dagger}+a_r\right)^{2m}|j\rangle
< \frac{(2m)!}{m!m!}\frac{(j+m)!}{j!}< 2^{2m}\frac{(j+m)!}{j!}
 \nonumber\\
&\frac{(j+m)!}{j!}&\sqrt{j+m+1}<
\langle j|\left(a_r^{\dagger}+a_r\right)^{2m+1}|j\pm 1\rangle
< \frac{(2m+1)!}{m!(m+1)!}\frac{(j+m)!}{j!}\sqrt{j+m+1}
< 2^{2m+1}\frac{(j+m)!}{j!}\sqrt{j+m+1}
\nonumber\\
\end{eqnarray}
we find that
\begin{equation}
\label{zeta}
\frac{|\zeta_{j,2m+1}|}{|\zeta_{j,2m}|}\sim 
\left(2\sqrt{\frac{1}{x_0^22\mu\nu_r}}\right)^{2m}\sqrt{j+m+1} 
\end{equation}
From (\ref{zeta}) we see that the series does not converge, as the term of the
expansion depends on the term $m$. However for a certain interval
corresponding to $j,m\le M_0$ such that
$\psi=\sqrt{1/x_0^22\mu\nu_r}\sqrt{M_0}\ll 1$, each term is bounded by the
corresponding term of a geometrical series with factor $\psi\ll 1$. For
typical values of a linear ion trap \cite{Linear} $\psi\sim\sqrt{M_0}/100$, so
that $M_0$ can assume very large values, $M_0 \ll
10^{4}$. \\
The divergence at orders $m>M_0$ is a signature of the divergence of the
Coulomb potential at $x=0$: such divergence requires the wavefunctions to be
zero in $x\le 0$, whereas the harmonic oscillator wavefunctions are different
from zero on all the space, although their occupation on the semiaxis $x\le 0$
is very small for $\psi\ll 1$. The divergence at $x=0$ is a mathematical
consequence, and does not correspond to the physical situation, in which the
ions move in a three dimensional space, although the confinement in the other
two dimensions is relatively tight. In principle one could build up a
potential that does not diverge in a specific point of the $x$ axis, whose
behaviour for $x>0$ tends to the Coulomb one. In this limit the series should
converge for low number states $j$. However such a detailed study is beyond
the scope of this paper \cite{Java2}.  We restrict to the case in which
$\psi\ll 1$, showing that in the chosen range the
harmonic approximation is quite good.\\
From this consideration it is then sufficient to compare the third order term
with the harmonic potential in order to show that the approximation is
sensible. The relation to be satified is then

\begin{equation}
\hbar\nu_r\gg \frac{e^2}{4\pi\epsilon_0 x_0^4}
\left(\sqrt{\frac{1}{2\mu\nu_r}}\right)^3 j^{\frac{3}{2}},
\end{equation}
with $j\le M_0$, which poses a further condition on $j,M_0$. Manipulating the
expression we find 

\begin{equation}
j\ll j_{\text{max}}\approx
\left(\frac{\hbar\nu_r}{e^2/4\pi\epsilon_0 x_0}\right)^{2/3}
\left(\frac{x_0}{\sqrt{\hbar/2\mu\nu_r}}\right)^{2}
\label{sfonda}
\end{equation}
which substantially agrees with the qualitative estimate in \cite{Manko}.
For $j\ll j_{\text{max}}$ the potential can be considered harmonic. For linear
traps \cite{Linear} 
$j_{\text{max}}\sim 120$, and the harmonic approximation is valid for
the region of energy we are considering. As an example the case $j=100$
corresponds to a correction of the order of $ 10^{-3}\hbar\nu_r$.

\section{Adiabatic elimination of the excited state}

We rewrite (\ref{Master}) in the following way
\begin{equation}
\frac{\text{d}}{\text{d}t}\rho=-\frac{i}{\hbar}[
H_{\text{eff}}\rho-\rho H_{\text{eff}}^{\dagger}]+J\rho,  
\label{re_Master}
\end{equation}
where $H_{\text{eff}}$ is the effective Hamiltonian
\begin{equation}
H_{\text{eff}}
=H_i+H_{\text{mec}}-i\frac{\gamma}{2} \sum_{i=1,2}\sigma_i^{\dagger}\sigma_i
\label{define0}
\end{equation}
and $J\rho$ the jump operator
\begin{equation}
J\rho
= \gamma \sum_{i=1,2}\sigma_i\tilde{\rho}\sigma_i^{\dagger}
\label{define1}
\end{equation}
We introduce the Liouvillians:

\begin{eqnarray}
L_0\rho&=&-\frac{i}{\hbar}[H_{\text{eff}}\rho-\rho
H_{\text{eff}}^{\dagger}]
\label{define_L0}\\
L_1\rho&=&- \frac{i}{\hbar}[V,\rho]\nonumber\\
L_2\rho&=&J\rho
\label{define2}
\end{eqnarray}
so that (\ref{re_Master}) can be rewritten as:

\begin{equation}
\frac{\text{d}}{\text{d}t}\rho=\left[L_0+L_1+L_2\right]\rho
\label{ME_2}
\end{equation}
In the limit $\Omega\ll \gamma$ we can eliminate the excited state in 
second order perturbation theory. Calling $P$ the projector onto the internal
ground state $|\text{g}\rangle$, and using a standard derivation based on 
projectors \cite{Gardiner}, we obtain the following equation for the ground
state of the system: 

\begin{eqnarray}
\frac{d}{dt}P\rho(t)
&=&PL_0P\rho(t)+\int_0^{t} d\tau
PL_1(1-P)\exp\left(L_0\tau\right)(1-P)L_1P\rho(t-\tau)\nonumber\\
&+&\int_0^{t} d\tau_1\int_{\tau_1}^{t} d\tau_2
PL_2(1-P)\exp\left(L_0\tau_1\right)(1-P)L_1\exp\left(L_0(\tau_2-\tau_1)\right)
P\rho(t-\tau_2)
\label{eq1}
\end{eqnarray}
where $P$ is a projector so defined on a density operator $X$
$PX=|\text{g}\rangle \langle \text{g}|X|\text{g}\rangle\langle\text{g}|$.
Markov approximation can be applied in the limit in which we may consider the
coupling to the excited state to evolve at a higher rate respect to the time
scale which characterizes the ground state evolution. 
This is true once we have moved to the interaction picture with respect
to the trap frequency. We define  

\begin{equation}
v_I(t)=\text{e}^{iH_{\text{mec}}t}P\rho(t)\text{e}^{-iH_{\text{mec}}t}
\end{equation}
and in the interaction picture (\ref{eq1}) has the form:

\begin{eqnarray}
&\frac{d}{dt}&v_I(t)
=\text{e}^{iH_{\text{mec}}t}\left[\int_0^{t} d\tau
PL_1\exp\left(L_0\tau\right)L_1\text{e}^{-iH_{\text{mec}}(t-\tau)}
v_I(t-\tau)\text{e}^{iH_{\text{mec}}(t-\tau)}\right]
\text{e}^{-iH_{\text{mec}}t}
\nonumber\\
&+           &\text{e}^{iH_{\text{mec}}t}
\left[\int_0^{t} d\tau_1\int_{\tau_1}^{t}d\tau_2 
PL_2\exp\left(L_0\tau_1\right)L_1\exp\left(L_0(\tau_2-\tau_1)\right)
CL_1\text{e}^{-iH_{\text{mec}}(t-\tau_2)}
v_I(t-\tau_2)\text{e}^{iH_{\text{mec}}(t-\tau_2)}\right]
\text{e}^{-iH_{\text{mec}}t}
\end{eqnarray}
Now we can neglect the change of $v_I$ during the time
$\tau$ on which the excited state evolves. Going back to the original 
reference frame we have now the equation in Markov approximation:

\begin{eqnarray}
\frac{d}{dt}P\rho(t)
&=&\int_0^{\infty} d\tau
PL_1\exp\left(L_0\tau\right)L_1\text{e}^{iH_0\tau}
P\rho(t)\text{e}^{-iH_0\tau}\nonumber\\
&+&\int_0^{\infty} d\tau_1\int_{\tau_1}^{\infty} d\tau_2
PL_2\exp\left(L_0\tau_1\right)L_1\exp\left(L_0(\tau_2-\tau_1)\right)
CL_1\text{e}^{iH_0\tau_2}P\rho(t)\text{e}^{-iH_0\tau_2}
\label{Markov}
\end{eqnarray}
We substitute now in (\ref{Markov}) the explicit form of the operators. 
After some algebra and application of the commutation rules
we obtain the following equation (where we have neglected the interference
terms between the two ions): 

\begin{eqnarray}
\frac{d}{dt}\rho(t)
&=     & PL_0\rho(t) - \sum_{i=1,2}\frac{\Omega_i^2}{4}
P\int_0^{\infty}\text{d}\tau\Bigl[\text{e}^{-(i\Delta+\gamma/2)\tau}
\text{e}^{ikx_i}\text{e}^{-iH_{\text{mec}}\tau}\text{e}^{-ikx_i}
\text{e}^{iH_{\text{mec}}\tau}\rho_1(t)
\nonumber\\
&+     &\text{e}^{-(-i\Delta+\gamma/2)\tau}
\rho_1(t)\text{e}^{-iH_{\text{mec}}\tau}\text{e}^{ikx_i}
\text{e}^{iH_{\text{mec}}\tau}\text{e}^{-ikx_i}\Bigr]
\nonumber\\
&+     & \sum_{i=1,2}\frac{\Omega_i^2}{4}
P\int_0^{\infty}\text{d}\tau_1\int_{\tau_1}^{\infty}\text{d}\tau_2
\int_{-1}^{1}\text{d}uN(u)\text{e}^{ikux_i}\nonumber\\
&\Bigl[&\text{e}^{-i\Delta(\tau_1-\tau_2)-\gamma/2(\tau_1+\tau_2)}
\text{e}^{-iH_{\text{mec}}\tau_1}
\text{e}^{-ikx_i}\text{e}^{iH_{\text{mec}}\tau_1}\rho_1(t)
\text{e}^{-iH_{\text{mec}}\tau_2}\text{e}^{ikx_i}\text{e}^{iH_{\text{mec}}
\tau_2} 
\nonumber\\
&+     &\text{e}^{+i\Delta(\tau_1-\tau_2)-\gamma/2(\tau_1+\tau_2)}
\text{e}^{-iH_{\text{mec}}\tau_2}\text{e}^{-ikx_i}
\text{e}^{iH_{\text{mec}}\tau_2}\rho_1(t)
\text{e}^{-iH_{\text{mec}}\tau_1}\text{e}^{ikx_i}
\text{e}^{iH_{\text{mec}}\tau_1} \Bigr]\text{e}^{-ikux_i}
\label{eq_final3}
\end{eqnarray}
Projecting (\ref{eq_final3}) on the basis of states $\left\{|{\bf
    n}\rangle\right\}$, we obtain Eq. (\ref{no_neglect}).

\begin{figure}
\begin{center}
\caption{Density of states $D(E)$ 
  plotted as a function of energy E in units of $\nu$. The grid is $\delta
  E=\nu/3$.}
\end{center}
\end{figure}

\begin{figure}
\begin{center}
\caption{(a) Spectrum of resonances for $\eta_0=0.1$ for a thermal
  distribution with average energy per mode $\bar{n}\nu=7.5\nu$, plotted on a
  grid of width $\nu/10$.  (b) Plot of $\langle n_0\rangle$ (solid line) and
  $\langle n_{r}\rangle$(dashed line) {\it vs}. time in units
  $t_F=2\gamma/\Omega^2$, for $\eta_0=0.1$, $\gamma=0.2\nu$,
  $\Omega=0.034\nu$, $\delta=-\nu$, and atoms initially in a flat distribution
  on the states with energy $E\le 15\nu$. (c) Population of COM mode $P_{n_0}$
  (onset) and of stretch mode $P_{n_r}$ (inset) {\it vs}. 
the respective vibrational
  state number at $t=600 t_F$.  }
\end{center}
\end{figure}

\begin{figure}
\begin{center}
\caption{(a) Comparison between rate equation (solid line) and QMC calculation
  (dashed line). Same parameters as in Fig. 2(b).  (b) Comparison between the
  time dependence of the COM average vibrational number as in Fig. 2(b) (solid
  line) and the average vibrational number for the case in which a single ion
  is cooled (dashed line).  For the single ion the mass has been rescaled so
  that $\eta^{(1)}=\eta_0$, $\nu^{(1)}=\nu$, $\gamma=0.2\nu$, 
  $\Omega=0.034\nu$, with an initially flat
  distribution for the first 15 states.}
\end{center}
\end{figure}

\begin{figure}
\begin{center}
\caption{(a) Spectrum of resonances for $\eta_0=0.6$ of ions in a thermal
  distribution with average energy per mode $\bar{n}\nu=7.5\nu$.  (b)
  Plot of $\langle n_0\rangle$ (solid line) and $\langle n_{r}\rangle$(dashed
  line) {\it vs}. time in unit $t_F=2\gamma/\Omega^2$, for $\eta_0=0.6$,
  $\gamma=0.2\nu$, $\Omega=0.034\nu$, $\delta=-2\nu$, and atoms initially in
  a flat distribution for states with energy $E\le 15\nu$. (c) Population
  of COM (onset) and of relative motion (inset) {\it vs}. 
the respective vibrational
  number state at $t=600 t_F$.  }
\end{center}
\end{figure}

\begin{center}
\begin{figure}
\caption{(a) Population $P_{\bf n}$ as a function of $n_0$ and $n_r$
  at a time $t=1000t_F$ and for $\gamma=0.02\nu$,
  $\Omega=0.17\gamma=0.0034\nu$, $t_F=2\gamma/\Omega^2=3460/\nu$. All the
  other physical parameters are the same as in Fig.  4(b),(c). (b),(c) Modulus
  square of the Franck Condon coefficients for the relative motion
  $F_{l,n_r}=|\langle l|\text{e}^{i\eta_r(a^{\dagger}_r+a_r)}|n_r\rangle|^2$ 
with $l=6,7$.  }
\end{figure}
\end{center}

\begin{center}
\begin{figure}
\caption{ 
Population $P_{\bf n}$ as a function of $n_0$ and $n_r$
at a time $t=600t_F$ and for $\gamma=\nu$, $\Omega=0.17\gamma=0.17\nu$,
$t_F=2\gamma/\Omega^2=69/\nu$. All the other physical parameters are the
same as in Fig. 4(b),(c).}
\end{figure}
\end{center}

\begin{center}
\begin{figure}
\caption{Time dependence of the average vibrational number of the COM mode for
  $\gamma=0.02\nu$ (dashed line), $\gamma=0.2\nu$ (solid line),
  $\gamma=0.4\nu$ (dashed-dotted line) and $\gamma=\nu$ (dotted line), keeping
  constant the ratio $\Omega/\gamma=0.17$. The time is in unit
  $t_F(\gamma)=2\gamma/\Omega^2\approx 70/\gamma$. All the other parameters
  are the same as in Fig. 4(b),(c).}
\end{figure}
\end{center}

\begin{figure}
\begin{center}
\caption{Comparison between rate equation (solid line) and QMC calculation
  (dashed line). Same physical parameters as in Fig. 4(b). The onset
  refers to the COM and the inset to the relative motion vibrational number.
}
\end{center}
\end{figure}

\begin{figure}
\begin{center}
\caption{Plot of the average vibrational number {\it vs}. 
time for the harmonic oscillator of
  frequency $\nu$ coupled to a second one with frequency 2$\nu$.  Comparison
  between rate equation (solid line) and QMC calculation (dashed line).
  Lamb--Dicke parameter for the mode $\nu$ is $\eta_{\nu}=0.6$,
  $\gamma=0.2\nu$, $\Omega=0.034\nu$, $\delta=-2\nu$, and atoms initially flat
  distributed on the states with energy $E\le 15\nu$.  }
\end{center}
\end{figure}

\begin{figure}
\begin{center}
\caption{
Plot of the average vibrational number of the COM mode {\it vs}. 
the time in unit $t_F$ for the case in which both ions are illuminated
(solid line, index 1) and only one ion is illuminated (dashed line). 
The solid line with index 2 corresponds to line 1 rescaled, 
where the time has been divided by a
factor 2. The other parameters are the same as in Fig. 4(b),(c).
} 
\end{center}
\end{figure}


\begin{references}

\bibitem{Reviews}
An overview of laser cooling can be found in S. Chu, Rev. Mod. Phys {\bf 70},
685 (1998); C. Cohen-Tannoudij, {\it ibidem} {\bf 70}, 707 (1998),
W.D. Phillips, {\it ibidem} {\bf 70}, 721 (1998). 

\bibitem{Salomon}
H. Perrin, A. Kuhn, I. Bouchoule and C. Salomon, Europhys. Lett. {\bf 42}, 395
(1998); S.E. Hamann, D.L. Haycock, G. Klose, P.H. Pax, I.H. Deutsch,
and P.S. Jessen, Phys. Rev. Lett. {\bf 80}, 4149 (1998).  

\bibitem{Lewenstein}
A theoretical study about possibilities of achieving Bose Einstein
condensation with laser cooling of
neutral atoms can be found in Y. Castin, J.I. Cirac and M. Lewenstein,
Phys. Rev. Lett. {\bf 80}, 5305 (1998) and references therein. 

\bibitem{BED}
A discussion about the advantages of using laser cooling for achieving a
quantum statistical regime can be found in J.I. Cirac,
M. Lewenstein and P. Zoller, Phys. Rev. Lett. {\bf 72}, 2977. 
(1994) and references therein. 

\bibitem{Zoller}
J.I. Cirac and P. Zoller, Phys. Rev. Lett. {\bf 74}, 4091 (1995).
For a review on ion trap quantum computer see 
A. Steane, Appl. Phys. B, {\bf 64} 623 (1997).



\bibitem{Linear}
M. Drewsen, C. Brodersen, L. Hornekaer, J.S. Hangst and J.P. Schiffer,
Phys. Rev. Lett. {\bf 81}, 2878 (1998); W. Alt, M. Block, P. Seibert and
G. Werth, Phys. Rev. A {\bf 58}, R23 (1998).
H.C. Naegerl, W. Bechter, J. Eschner, F. Schmidt-Kaler and R. Blatt,
Appl. Phys. B, {\bf }  (1998). 


\bibitem{Stenholm}  S.~Stenholm, Rev. Mod. Phys. {\bf 58} , 699 (1986).

\bibitem{Monroe}
F.~Diedrich, J.C.~Berquist, W.M.~Itano, and D.J.~Wineland,
Phys. Rev. Lett. {\bf 62}, 403 (1989); C.~Monroe, D.M.~Meekhof,
B.E.~King, S.R.~Jefferts, W.M.~Itano, D.J.~Wineland, and P.~Gould,
Phys. Rev. Lett. {\bf 75}, 4011 (1995). 


\bibitem{Irene}
I. Marzoli, J.I. Cirac, R. Blatt and P.Zoller, Phys. Rev. A, {\bf 49} 2771
(1994).

\bibitem{Wineland}
B.E. King, C.S. Wood, C.J.Myatt, Q.A. Turchette, D. Leibfried, W.M. Itano,
C. Monroe and D.J. Wineland, Phy. Rev. Lett. {\bf 81}, 1525 (1998).


\bibitem{noLDLio}
G. Morigi, J.I. Cirac, M. Lewenstein and P. Zoller, Europhys. Lett. {\bf 39}
13 (1997).

\bibitem{noLDL}
G. Morigi,  J.I. Cirac, K. Ellinger and P. Zoller, Phys. Rev. A
{\bf 57}, 2909 (1998); D. Stevens, A. Brochards and A.M. Steane, Phys. Rev. A
{\bf 58}, 2750 (1998); L. Santos and M. Lewenstein, Report No. 
quant-ph/9808014.

\bibitem{Vogt}
A.W. Vogt, J.I. Cirac and P. Zoller, Phys. Rev. A, {\bf 53} 950 (1996).

\bibitem{Java}
J. Javanainen, J. Opt. Soc. Am. B, {\bf 5}, 73 (1988). 


\bibitem{clusters}
J.N. Tan, J.J. Bollinger, B. Jelenkovic, and D.J. Wineland, Phys. Rev. Lett.
{\bf 75}, 4198 (1995); F. Diedrich, E. Peik, J.M. Chen, W. Quint and
H. Walther, {\it ibidem} {\bf 59}, 2935 (1987); 
W.M. Itano, J.J. Bollinger, J.N. Tan, B. Jelenkovic, X.-P. Huang, and D.J.
Wineland, Science {\bf 279}, 686 (1998); G. Birkl, S. Kassner and H. Walther,
Nature {\bf 357}, 310 (1992). 

\bibitem{footnote1} 
  The final energy obtained by means of Doppler cooling of
  single atoms in a trap depends on the internal transition quantum numbers,
  and for hyperfine states with selection rule $\Delta m_F=\pm 1$ it is
  $\langle n\rangle\nu = 7\gamma/20$; see \cite{Stenholm}.


\bibitem{James}
D.F.V. James, Appl. Phys. B, {\bf 66}, 181 (1998).

\bibitem{Lange} W. Lange, private communication.

\bibitem{Parkins}
R. Blatt, J.I. Cirac and P.Zoller, Phys. Rev. A, {\it 52}, 518 (1995).
J.I.~Cirac, A.S.~Parkins,
R.~Blatt, P.~Zoller, Adv. At. Mol. Phys., ed B.~Bederson and
H.~Walther, {\bf 37}, 237 (1996).   

\bibitem{footnote2}
This argument is valid in the limit of low saturation. At saturation
the antibunched nature of the atoms' spontaneous emission plays a role by
modifying the diffusion. See \cite{Stenholm}.

\bibitem{footnote5}
Phenomena due to interference between the internal paths of excitation, which
lead to superradiance and subradiance effects, occur for distances between
the ions of the order of the wavelength, when dipole--dipole interaction
cannot be neglected. See R.G. DeVoe and R.G. Brewer, Phys. Rev. Lett. {\bf
  76}, 2049 (1996); R.G. Brewer, {\it ibidem} {\bf 77}, 5153 (1996) and 
\cite{Vogt}.  


\bibitem{Manko}
V.V. Dodonov, V.I. Manko and L. Rosa, Phys. Rev. A {\bf 57}, 2851 (1998)

\bibitem{Java2} A detailed discussion about the approximation may be found in
  J. Yin and J. Javanainen, Phys. Rev. A {\bf 51}, 3959 (1995).

\bibitem{Gardiner}
C.W. Gardiner, {\it Quantum Noise}, Springer Verlag (1991).



\end{references}
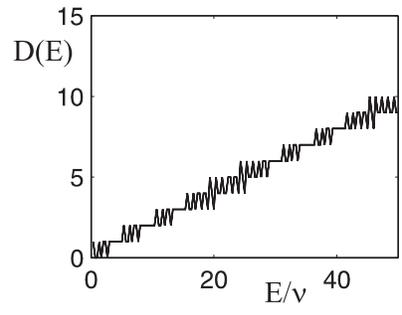
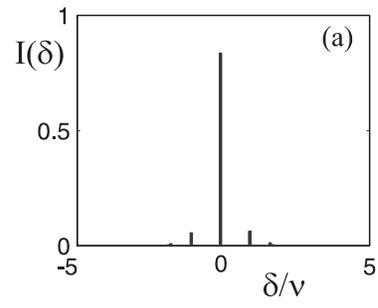
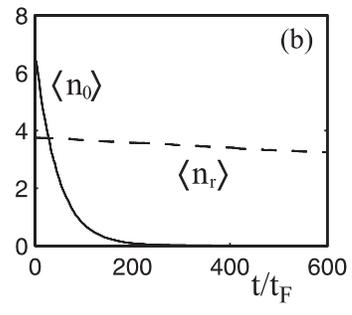
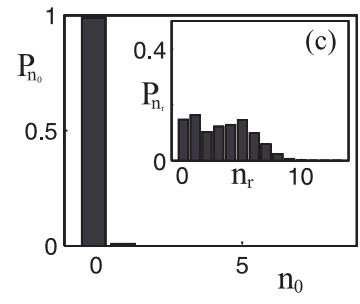
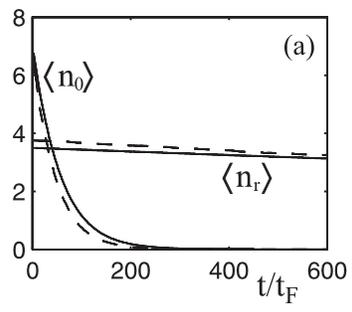
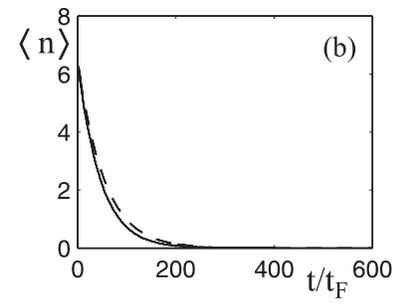
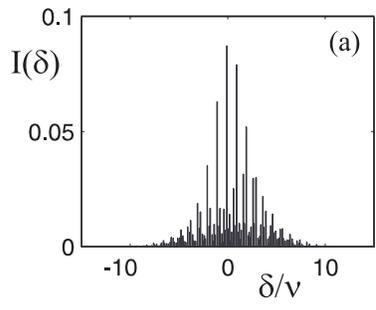
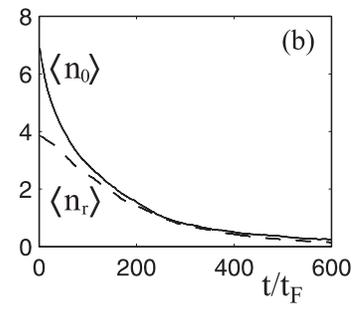
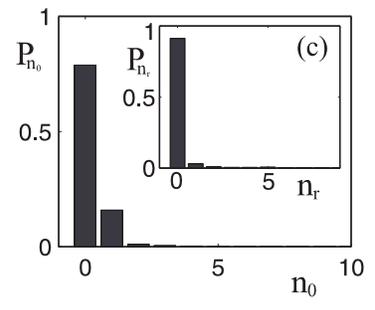
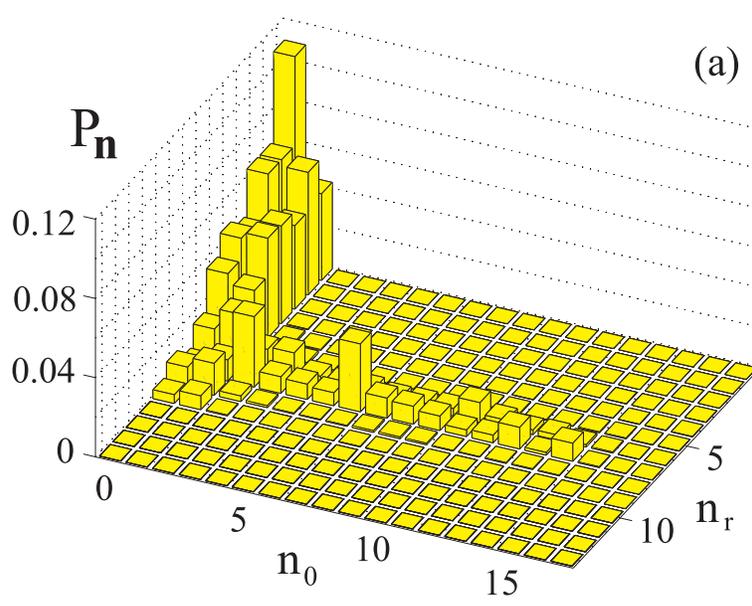
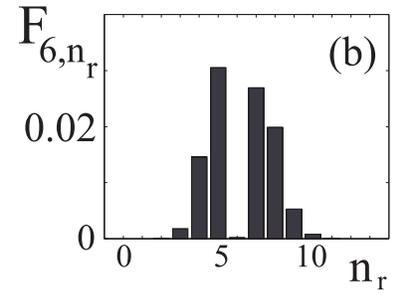
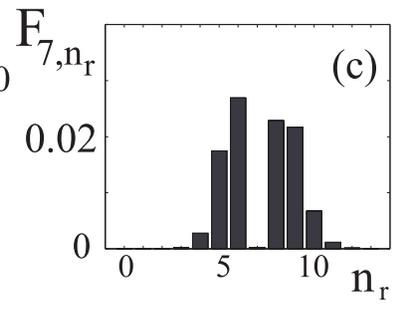
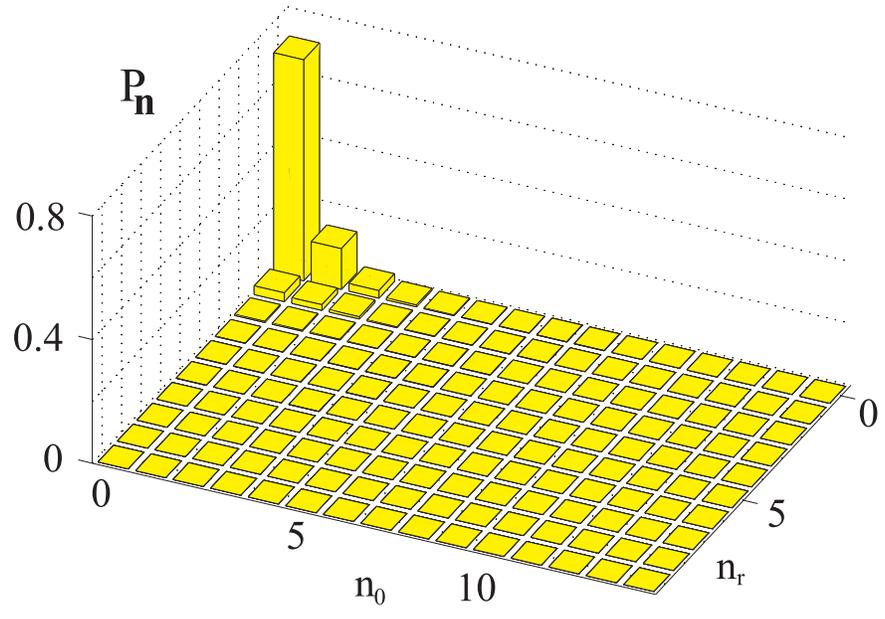
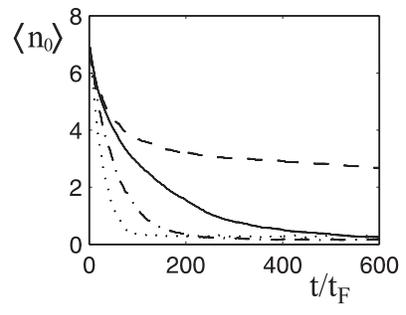
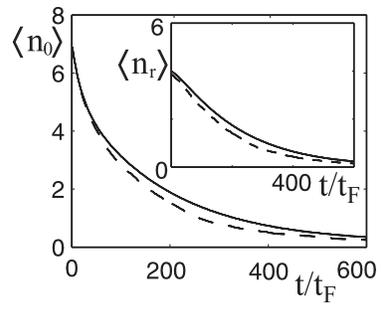
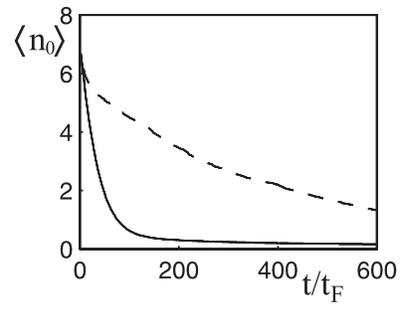
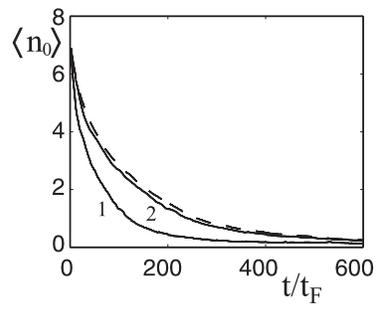
\end{document}